%% file: main.tex
\documentclass[%
 aip,
 amsmath,amssymb,
preprint,
]{revtex4-1}

\usepackage{dcolumn}
\usepackage[mathlines]{lineno}

\usepackage[utf8]{inputenc}
\usepackage[T1]{fontenc}
\usepackage{mathptmx}

\usepackage{hyperref}
\usepackage{latexsym}
\usepackage{mathrsfs}
\usepackage{graphicx}
\usepackage{subfig}
\usepackage{bm}
\usepackage[dvipsnames]{xcolor}
\usepackage[section]{placeins}
\usepackage{amsmath}
\usepackage[normalem]{ulem}
\usepackage[titletoc,title]{appendix}

\usepackage{iftex}

\usepackage{pgfplots}
\usepackage{pgfplotstable}

\include{Commands}

\include{CommandsFigures}

\begin{document}

\title[]{Nonlinear generation of global zonal structures in gyrokinetic simulations of TCV and ASDEX Upgrade magnetic configurations.}

\author{I. Novikau}
\affiliation{Max-Planck-Institut f\"ur Plasmaphysik, 85748 Garching, Germany}

\author{A. Biancalani}
\email{alessandro.biancalani@devinci.fr}
\affiliation{De Vinci Higher Education, De Vinci Research Center, 92 916 Paris, France}

\author{A. Bottino}
\author{E. Poli}
\author{G. D. Conway}
\affiliation{Max-Planck-Institut f\"ur Plasmaphysik, 85748 Garching, Germany}

\author{P. Manz}
\affiliation{Institute of Physics, University of Greifswald, 17489 Greifswald, Germany}

\author{L. Villard}
\author{N. Ohana}
\affiliation{{\'E}cole Polytechnique F\'ed\'erale de Lausanne, Swiss Plasma Center, CH-1015 Lausanne, Switzerland}

\author{ASDEX Upgrade Team\footnote{See author list of 
\href{https://doi.org/10.1088/1741-4326/ad249d}{H. Zohm \textit{et al 2024 Nucl. Fusion}, \textbf{64} 112001} for the ASDEX Upgrade Team.}}
\affiliation{Max-Planck-Institut f\"ur Plasmaphysik, 85748 Garching, Germany}

\date{\today}

\begin{abstract}
In tokamaks, turbulence is responsible not only for the anomalous transport of heat and particles from the core to the edge, which reduces heating efficiency, but also for the generation of zonal structures (ZSs). Examples of ZSs are those with characteristic sound frequency, like the geodesic acoustic modes (GAMs).
Developing a theoretical model of ZS is essential, as they contribute to the turbulence saturation and thus indirectly influence transport. 
In this paper, we investigate the radial structure of ZS in the frequency range of GAMs, by means of numerical simulations using the gyrokinetic particle-in-cell code ORB5. 
We find that radially extended coherent ZSs (labelled here as global ZSs) are nonlinearly generated by the high-$n$ part of the turbulence spectrum (with $n$ being the toroidal mode number) by means of self-consistent simulations. 
We also reproduce this generation by mimicking the turbulence modes with an antenna, thus isolating the nonlinear generation mechanism.
\end{abstract}

\maketitle
\input{Introduction}

\input{theory}
\input{sim_without_ant}
\input{antenna}

\input{Conclusions}

\begin{acknowledgments}
Discussions with \"O. G\"urcan and P. Morel are gratefully acknowledged.
This work has been carried out within the framework of the EUROfusion Consortium and has received funding from the Euratom research and training program 2014-2018 and 2019-2020 under grant agreement N$^o$ 633053. 
Also, this work has been carried out within the framework of the EUROfusion Consortium, partially funded by the European Union via the Euratom Research and Training Programme (Grant Agreement N$^o$ 101052200 — EUROfusion). 
The Swiss contribution to this work has been funded by the Swiss State Secretariat for Education, Research and Innovation (SERI). 
This work was supported in part by the Swiss National Science Foundation.
Views and opinions expressed are however those of the author(s) only and do not necessarily reflect those of the European Union, the European Commission or SERI. 
Neither the European Union nor the European Commission nor SERI can be held responsible for them.
Simulations, presented in this work, have been performed on the CINECA Marconi supercomputer within the framework of the OrbZONE and ORBFAST projects.
\end{acknowledgments}

\bibliography{./main.bib}

\end{document}

%% file: Commands.tex
\newcommand{\aeq}{\begin{equation}}
\newcommand{\eeq}{\end{equation}}
\newcommand{\aeqn}{\begin{eqnarray}}
\newcommand{\eeqn}{\end{eqnarray}}
\newcommand{\aeqns}{\begin{eqnarray*}}
\newcommand{\eeqns}{\end{eqnarray*}}
\newcommand{\yb}[1]{\mathbf{#1}}
\newcommand{\ybs}[1]{\boldsymbol{#1}}

\newcommand*\diff{\mathop{}\!\mathrm{d}}

\newcommand{\pa}{\chi}
\newcommand{\ta}{\varphi}



%% file: Introduction.tex
\section{Introduction}
\label{sec:introduction}

Gyrokinetic (GK) simulations have become one of the most prominent and widely used numerical tools in the magnetic fusion plasma community.
They can favor the development of analytical theories by providing additional amount of data about plasma behavior in different geometries and with various equilibrium profiles. 
Due to their capability to model plasma systems including realistic geometric effects and wave-particle interactions, GK models serve as one of the dominant numerical engines for comparison with and explanation of experimental plasma data. 

Drift instabilities, such as Ion Temperature Gradient (ITG) modes~\cite{Rudakov61} that naturally occur in tokamak plasmas due to finite ion temperature gradients, can drive the formation of zonal structures (ZSs)~\cite{Hasegawa79,Diamond91, Lin98}, whose electric field is toroidally and mainly poloidally symmetric and has a finite radial wavenumber. 
The oscillatory branch of these structures is so-called Geodesic Acoustic Modes (GAMs)~\cite{Winsor68, Conway22, Zhou25}. Modes with frequency on the range of GAMs can be observed in tokamaks with continuum or global frequency spectra.
The continuum GAM branch is characterized by a local frequency with a clear dependence on plasma temperature and safety factor radial profiles~\cite{Jakubowski02, Hamada05, Conway05, McKee06, Zonca08, Vermare12}. 
The global branch exhibits radially elongated ZS with a spatially uniform frequency.
Such spectra have been observed both experimentally~\cite{Conway08, Kong13, Wang13, Meijere14, Huang18, McCarthy25} and in nonlinear numerical simulations~\cite{Merlo18, Villard19}. 
Based on a detailed numerical investigation of the linear properties of GAMs, the existence of these global ZS has been conjectured to originate from nonlinear effects~\cite{Novikau17}.

In this work, we investigate the nonlinear interaction of ITGs and ZSs in two realistic magnetic configurations: one of the TCV machine, which was considered in~\onlinecite{Villard19}, and one of the ASDEX Upgrade (AUG) tokamak (discharge \#20787) described in~\onlinecite{Conway08} and considered in~\onlinecite{Novikau17}.
In particular, in this paper, we analyze the zonal mode generation by different ITG spectra by performing nonlinear simulations with the global GK particle-in-cell code ORB5~\cite{Jolliet07, Bottino15, Lanti19}. 
ORB5 has been originally developed for turbulence studies, and then it has also studied ZSs with a detailed phase of verification against analytical theory and benchmark against other GK codes~\cite{Biancalani17}.
We show that high toroidal mode numbers are responsible for the excitation of the global ZSs, and this tendency is observed in both configurations. 

Being a complex system, a plasma has a variety of waves and instabilities of diverse time and space scales. 
The modes are often coupled to each other. 
Interaction of the ZS with the ITG instabilities is a typical example of the complex interconnection.
The ITG modes arise because of nonzero background ion temperature gradients, while the ZSs being excited by the ITGs lead to the saturation of the non-zonal modes that slows down the ZS development as well\cite{Naulin05, Scott05, Medvedeva17, Hahm24, Feng25, Albert25}.
To partly disentangle the coupling between these modes, one can replace one of the field components (zonal or non-zonal one) by an externally applied field perturbation, the so-called antenna.
Such kind of technique was implemented for the first time in ORB5 in~\onlinecite{Ohana18}. 
There, the antenna had a stationary ZS, and its effect on the dynamics of (non-zonal) ITGs was studied.
This work was extended later on in Refs.~\onlinecite{Ohana19, OhanaThesis20}, where the stationary zonal antenna was replaced by an oscillatory sheared ZS.
It was demonstrated that the antenna with a modest frequency, changes the ITG stabilization effect in time, 
while by increasing the ZS frequency the stabilization effect decreases according to an effective reduced shearing rate, in agreement with Ref.~\onlinecite{Hahm99}.
Moreover, in~\onlinecite{OhanaThesis20}, early tests with the antenna mimicking a slab-like drift wave were performed.

The goal of the present paper is to complete the previous preparatory work, by using the antenna to mimic (non-zonal) ITGs with realistic spatial structures and study the nonlinear generation of global ZSs.
By performing numerical simulations with a broad ITG spectrum, one can afterwards extract space structure of a chosen ITG mode with a certain toroidal mode number $n$.
In such a way, one can fix the ITG mode structure and use it as an antenna with a certain oscillation frequency.
Finally, by imposing such kind of externally applied electrostatic field in the TCV plasma, we show that we can reproduce the formation of the global ZS.
In the AUG plasma, GAMs generation by antenna is observed as well, and the zonal mode spectrum is recreated.

The paper is divided in the following sections. In Sec.~\ref{sec:ant-implementation}, the non-zonal antenna is defined as an external field added to the particle trajectories, and its discretization and implementation in ORB5 is described.  Sec.~\ref{sec:tcv-aug-original} shows the results of self-consistent simulations where the ZSs are nonlinearly generated by ITG turbulence in TCV and AUG tokamaks, in the absence of an imposed antenna. In this section, in particular, the evidence of globally extended ZSs is shown, and the part of the ITG spectrum responsible for their generation is identified. Then, in Sec.~\ref{sec:tcv-aug-antenna}, we mimic the ITG modes with an antenna and use it to artificially generate the global ZSs. With this method, we can isolate and study the nonlinear generation mechanism, and identify the threshold of continuum/global behavior. Finally, Sec.~\ref{sec:conclusions} is dedicated to a discussion on the results, and an outlook on the future steps.

%% file: theory.tex
\section{Antenna implementation}
\label{sec:ant-implementation}

\subsection{Theory}
Theoretical aspects of the antenna realization has been described in~\onlinecite{Ohana18}, and most recently in~\onlinecite{OhanaThesis20}. 
Antenna is represented as an additional electrostatic (ES) potential $\Phi_a$, which amplitude is considered one order smaller than that of background quantities.
This potential perturbs particle characteristics as
\begin{eqnarray}
&&\yb{\dot R} = \yb{\dot R}_{\rm orig} - \frac{c \ybs{\nabla}(J_0\Phi_a) \times \yb{b}}{B^*_\parallel},\label{eq:R-ant}\\
&&\dot p_z = \dot p_{z, {\rm orig}} - \frac{Z e m}{p_z}(\yb{v}_\parallel + \yb{v}_{\nabla p} + \yb{v}_{curvB})
\cdot \ybs{\nabla}(J_0 \Phi_a),\label{eq:pz-ant}
\end{eqnarray}
where $\yb{\dot R}$, $\dot p_z$ are marker equations of motion of the real space coordinate $\yb{R}$ and parallel canonical momentum $p_z$, and $\yb{\dot R}_{\rm orig}$, $\dot p_{z, {\rm orig}}$ are the corresponding characteristics in an electrostatic GK system without antenna:
\begin{eqnarray}
&&\yb{\dot{R}}_{\rm orig} = \ybs{v}_\parallel + \ybs{v}_{\nabla B} +  
\ybs{v}_{curvB} + \ybs{v}_{\nabla p} + 
\ybs{v}_{E\times B},
\label{eq:R-v-terms}\\
&&\dot{p}_{z,{\rm orig}} = - \frac{m}{p_z} 
(\ybs{v}_\parallel + \ybs{v}_{\nabla p} + \ybs{v}_{curvB})\cdot 
\ybs{\nabla}(\mu B + Ze(J_0\Phi)),
\label{eq:p-v-terms}\\
&&\ybs{v}_{\parallel} = \frac{p_z}{m}\ybs{b},
\label{eq:vpar}\\
&&\ybs{v}_{\nabla B} = \frac{c\mu B}{Z e B^*_\parallel} 
\ybs{b} \times \frac{\ybs{\nabla} B}{B},
\label{eq:vbnB}\\
&&\ybs{v}_{curvB} = \left(\frac{p_z}{m}\right)^2 
\frac{cm}{Z e B^*_\parallel}
\ybs{b} \times \frac{\ybs{\nabla} B}{B},
\label{eq:vcB}\\
&&\ybs{v}_{\nabla p} = - \left(\frac{p_z}{m}\right)^2 
\frac{cm}{Ze B^*_\parallel} \ybs{b} \times 
\left(\ybs{b} \times \frac{\ybs{\nabla}\times 
	\yb{B}}{B} \right),\label{eq:vgradp}\\
&&\ybs{v}_{E\times B} = - \frac{c\ybs{\nabla}(J_0\Phi)
	\times \ybs{b}}{B^*_\parallel}.\label{eq:vexb}
\end{eqnarray}
Here, $Ze$, $m$ are particle charge ($Z_e = -1$) and mass, $c$ is the speed of light, $\mu$ is the magnetic moment, $\yb{b}=\ybs{B}/B$, $B^*_\parallel = \ybs{b}\cdot \ybs{B}^*$ with $\ybs{B}^* = \ybs{B} + (p_z c/Ze)\ybs{\nabla}\times\ybs{b}$, where $\ybs{B}$ is the background magnetic field,
$J_0$ is the gyroaveraging operator defined as 
$(J_0\psi) ( \mathbf{R},\mu) = (1/2\pi)\int_0^{2\pi} \psi( \mathbf{R}+\boldsymbol{\rho}(\theta)) \diff \theta$ being acting on an arbitrary function $\psi$.
The ES potential $\Phi$ (so-called self-consistent field) arises from the GK Poisson equation due to plasma density perturbation.
In linear simulations, the term~\ref{eq:vexb} is omitted. However, the antenna contribution to Eqs.~\ref{eq:R-ant} and~\ref{eq:pz-ant} are kept in both linear and nonlinear calculations.
Finally, although the antenna modifies particle trajectories, the corresponding plasma perturbation does not influence the antenna field (which is not the case of the self-consistent field) since the $\Phi_a$ is not calculated by the Poisson equation, but is defined by a user at the beginning of a simulation as it is described in Section~\ref{eq:num-real}. 

\subsection{Numerical realization}
\label{eq:num-real}
ORB5 is a particle-in-cell (PIC) code, where a particle-based algorithm for the Vlasov equation is coupled with a grid-based technique for the modeling of the self-consistent electromagnetic fields~\cite{Lanti19, OhanaThesis20}.
Plasma particle dynamics is described by the so-called markers that follow the GK equations of motion (the so-called characteristics), and the time evolution of the marker weights satisfies the GK Vlasov equation~\cite{Jolliet07, Bottino15, Tronko16, Lanti19}.
On the other hand, the finite element representation of the electromagnetic potentials with the B-spline basis functions is employed in the code to describe the field dynamics.
Since the current version of the antenna can be used only in electrostatic simulations, the magnetic potential dynamics is not considered in this work.
The electric potential is discretized with three-dimensional finite elements expressed as the triple product of one-dimensional components 
\begin{equation}
\Phi(s,\pa,\ta) = \sum_{i = 0}^{N_s + p - 1}\sum_{j = 0}^{N_\pa + p - 1}\sum_{k = 0}^{N_\ta + p - 1} 
\widehat\Phi_{ijk}\Lambda^{(p)}_{i}(s)\Lambda^{(p)}_{j}(\pa)\Lambda^{(p)}_{k}(\ta).\label{eq:FE-DFT-to-Real}
\end{equation}
Here, $\Phi(s,\pa,\ta)$ is the ES potential in the real space with $s = \sqrt{\psi/\psi_{edge}}$ being the radial coordinate (where $\psi$ is the poloidal flux coordinate), $\pa$ is the so-called straight-field-line coordinate ~\cite{Lanti19, DHaeseleer1991}, $\ta$ is the toroidal angle; $\widehat\Phi_{ijk}$ are some real values.
The finite elements $\Lambda^{(p)}_l$ are expressed by B-splines, which in ORB5 can be up to the third order ($p = 3$).
Due to this representation, the GK Poisson equation derived from a variational principle~\cite{Tronko16} can take a form of the following system of equations:
\begin{equation}
\sum_{i^\prime j^\prime k^\prime} Q_{ijk, i^\prime j^\prime k^\prime}\widehat\Phi_{i^\prime j^\prime k^\prime} = 
\widehat\rho_{ijk}.\label{eq:discr-poisson}
\end{equation}
Every $Q_{ijk, i^\prime j^\prime k^\prime}$ element is calculated as a sum of several terms in the GK Poisson equation,
including different geometric coefficients of a considered magnetic field configuration.
The $\widehat\rho_{ijk}$ elements on the right-hand-side represent projection of the marker weights on the field grid.
The density and potential elements can be discrete Fourier transformed (DFT) in the poloidal and toroidal directions as
\begin{equation}
\widetilde{\widehat f}_{imn} = \sum_{j=0}^{N_\pa - 1}\sum_{k=0}^{N_\ta - 1}\widehat f_{ijk} 
e^{-2\pi i \left(mj/N_\pa + nk/N_\ta \right)},\label{eq:DFR}
\end{equation}
where $\widehat f_{ijk}$ can be density $\widehat \rho_{ijk}$ or $\widehat \Phi_{ijk}$ elements. Here, the poloidal $m$ and toroidal $n$ numbers are introduced.
To speedup simulations and to keep only the most pronounced modes, the code performs the field-aligned filtering on the density Fourier elements $\widetilde{\widehat \rho}$, where it simulates only the poloidal modes $m$ that satisfy the condition $|m + n q(s)| < \Delta m$ with $\Delta m$ being set by a user and being usually around $5$, while $q$ is the safety factor profile defined by a background magnetic field configuration. 
This means that for a fixed $n$ and radial position $s$ the code simulate only $2\cdot\Delta m + 1$ poloidal modes.
Apart from that, only the toroidal modes within the interval $[n_{min}, n_{min} + n_{step},..., n_{max}]$ with some step $n_{step} \geq 1$ are modeled. 
The highest simulated toroidal mode $n_{highest}$ is defined from the $\mod(n_{highest} - n_{min}, n_{step}) = 0$ and $n_{highest} \leq n_{max}$ conditions, where $n_{min}, n_{max}, n_{step}$ are input parameters.
One should note here as well that due to the realness of the density and potential elements, their Fourier spectrum is symmetric, and one can calculate only half of the Fourier coefficients ($n > 0$).
The rest of the coefficients can be found as $\widetilde{\widehat f}_{i,-m,-n} = \widetilde{\widehat f}_{imn}$.

Using the described discrete Fourier transform and the filtering, one can solve the GK Poisson equation~\ref{eq:discr-poisson} directly in the Fourier space
\begin{eqnarray}
&&\sum_{i^\prime = 0}^{N_s + p - 1} \sum_{n^\prime = n_{min}}^{n_{max}} 
\sum_{m^\prime = - n^\prime q_{i_c} - \Delta m}^{- n^\prime q_{i_c} + \Delta m} 
\widetilde Q_{imn,i^\prime m^\prime n^\prime} \widetilde{\widehat \Phi}_{i^\prime m^\prime n^\prime} = 
\widetilde{\widehat \rho}_{imn},\label{eq:discr-poisson-fourier}\\
&&q_{ic} = q\left(\frac{i - (p-1)}{2}\Delta s\right) 
\end{eqnarray}
where $\Delta s$ is the step of the radial grid, and the $m, n$ mode numbers have to satisfy the filter described earlier.
The $\widetilde Q_{imn,i^\prime m^\prime n^\prime}$ elements can be calculated from $Q_{ijk, i^\prime j^\prime k^\prime}$ (corresponding details can be found in~\onlinecite{OhanaThesis20}, Chapter 3). 

To create an ITG antenna with a ballooning space structure at a given time moment, one chooses a toroidal mode number of interest $n_a$ and save the corresponding DFT of the potential finite elements $\widetilde{\widehat \Phi}_{imn_a}$ for all radial points $i$ and poloidal numbers $m$ allowed by the filter described after Eq.~\ref{eq:DFR}. 
In such a way, one stores a whole radial and poloidal structure of the chosen toroidal mode $n_a$ that is originally generated in some preliminary GK simulation.
It is worth noting that it would be unreasonable to use the field representation in real space to store a realistic structure of electrostatic potential in real coordinates. 
As it was alluded earlier, since the code operates with discrete Fourier representation of the finite elements (Eq.~\ref{eq:DFR}), one can save the field structure by using much smaller amount of memory.
The number of poloidal and toroidal modes can significantly smaller than that of the poloidal and toroidal grid points due to the field-aligned filtering.

To describe the rotation of the $n_a$ antenna, one can use the procedure described in Sec.~\ref{sec:ant-time-evol}, where one has to save just two snapshots of the antenna at two different time moments.
The stored $\widetilde{\widehat \Phi}_{imn_a}$ antenna finite elements can be applied later to another simulation as an external source, which space structure and frequency do not depend anymore on the plasma evolution.
The $\widetilde{\widehat \Phi}_{imn_a}$ elements are transformed by the code back to the real coordinates using Eq.~\ref{eq:FE-DFT-to-Real}.
The resulting antenna electric potential is applied finally to the marker characteristics as it is shown in Eqs.~\ref{eq:R-ant} and~\ref{eq:pz-ant}.

\subsection{Antenna time evolution}
\label{sec:ant-time-evol}
To describe the antenna evolution in time, one can save the antenna space structure at two time moments with a quarter-period interval. Indeed, by representing for the sake of simplicity the antenna at a radial point $s_1$ as
\aeqn
\Phi_a = A_a \cos(\phi n_1 + m\theta - \omega_a t)
\eeqn
with a constant amplitude $A_a$ and frequency $\omega_a$, one can directly rewrite the expression as  
\aeqn
\Phi_a = A_a \cos(\phi n_a + m\theta) \cos(\omega_a t) - A_a \cos(\phi n_a + m\theta + \pi/2) \sin(\omega_a t).
\label{eq:ant-tim-evol}
\eeqn
The $\cos(\phi n_a + m\theta)$ is equivalent to $\Phi_a(n_a, m, s_1, t_0)$, while 
$\cos(\phi n_a + m\theta + \pi/2)$ is equivalent to the potential space structure a quarter-period earlier $\Phi_a(n_a, m, s_1, t_0 - 0.25 T_a)$.
In other words, the antenna time evolution is described just by its frequency $\omega_a$, and its space structures at two time moments.
In this case, one obtains a purely oscillating ES potential with the same frequency for every radial point and for every toroidal and poloidal mode numbers.

%% file: sim_without_ant.tex
\section{Global ZS excitation by self-consistent ITG modes}
\label{sec:tcv-aug-original}

\begin{figure}[!t]  
    \subfloat{\includegraphics[width=0.38\textwidth]{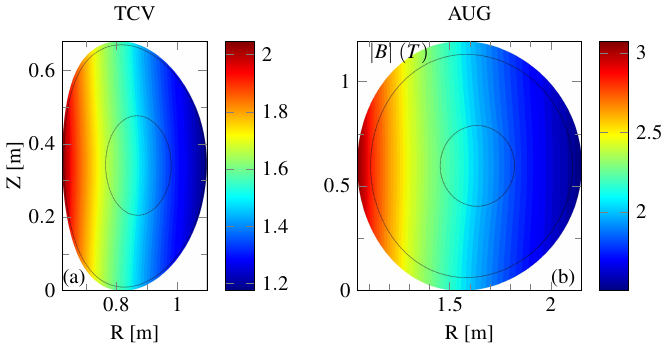}}
    \subfloat{\includegraphics[width=0.60\textwidth]{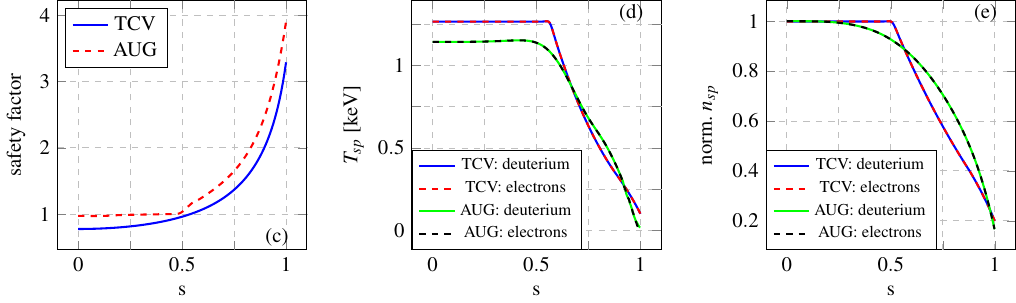}}
    \caption{\label{fig:TCV-AUG-conf}
        Magnetic field configuration ($|B|(T)$) in the TCV (a) and ASDEX Upgrade (b) cases. 
            Safety factor
        (c): Safety factor profiles. (d): Temperature profiles. (e): Density profiles.
        TCV: $R_0 (m) = 0.88$, $a_0 (m) = 0.25$, $B_0 (T) = 1.44$, $\kappa = 1.44$, $L_x = 265.5$ ($\rho^* = 7.53\cdot 10^{-3}$).
        AUG: $R_0 (m) = 1.65$, $a_0 (m) = 0.50$, $B_0 (T) = 2.0$, $\kappa = 1.08$, $L_x = 1077.0$ ($\rho^* = 1.85\cdot 10^{-3}$).
        \label{eq:BTn}
    }
\end{figure}
\begin{figure}[!t]  
    \includegraphics[width=0.99\textwidth]{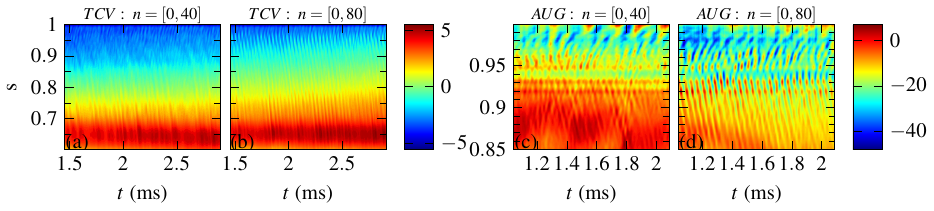}
    \caption{\label{fig:orig-erbar-st}
            Evolution in time of the zonal radial electric field, $\overline{E}_r[a.u]$, for the TCV (a,b) and the AUG (c,d) configurations in simulations with different ITG spectra.
    }
\end{figure}

In this section, we perform nonlinear GK simulations of the TCV and AUG magnetic configurations, Fig.~\ref{fig:TCV-AUG-conf}, reconstructed from experimental data using the Grad-Shafranov solver code CHEASE~\cite{Lutjens96}.
The TCV temperature radial profiles are taken from~\cite{Villard19}.
They are analytically created for a qualitative investigation of the global zonal structures.
The AUG profiles are taken from the AUG discharge \#20787 from~\cite{Conway08}, where a staircase spectrum of modes in the frequency range of GAMs was observed.
For the sake of normalization, we use $s_0 = 0.88$, where $T_e(s_0)\ [keV] = 0.343$, as a reference flux-surface (radial position) in the TCV case, while $s_0 = 0.95$ ($T_e(s_0)\ [keV] = 0.165$) in the AUG modeling.
The modeling is performed electrostatically with adiabatic electrons and gyrokinetic thermal ions (deuterium).

To understand which part of the ITG spectrum is responsible for the global ZS formation, nonlinear simulations with different ITG toroidal spectra are launched.
For example, simulations with maximum toroidal mode number $n_{max} = 80$ have the following parameters
\begin{eqnarray}
&&TCV:\ n = [0, 80],\  m = [-270, 270],\ n_\phi = 320,\ n_\chi = 640,\  n_s = 256,\ N_d = 6\cdot 10^8,\\
&&AUG:\ n = [0, 80],\  m = [-325, 325],\ n_\phi = 336,\ n_\chi = 672,\  n_s = 256,\ N_d = 4.8\cdot 10^8,
\end{eqnarray}
where $n_\phi$ is the number of grid points in the toroidal direction, $n_\chi$ is the number of the poloidal grid points, $N_d$ is the number of the deuterium markers.
Maximum allowed poloidal number is defined by the filter described in~\ref{sec:ant-implementation}.
The minimum amount of points in the toroidal and poloidal grids is $4$ per toroidal mode and $2$ per poloidal mode. 
The TCV case is calculated in $s = [0.5, 1.0]$ radial domain chosen to keep only the area with a non-zero temperature gradient, Fig.~\ref{fig:TCV-AUG-conf}d. 
The number of points in the radial grid is $n_s = 256$.
The AUG configuration is simulated in $s = [0.4, 1.0]$ radial domain with $n_s = 1200$. The time step is $dt[\omega_{ci}^{-1}] = 20$.

\begin{figure}[!t]  
    \subfloat{\includegraphics[width=0.99\textwidth]{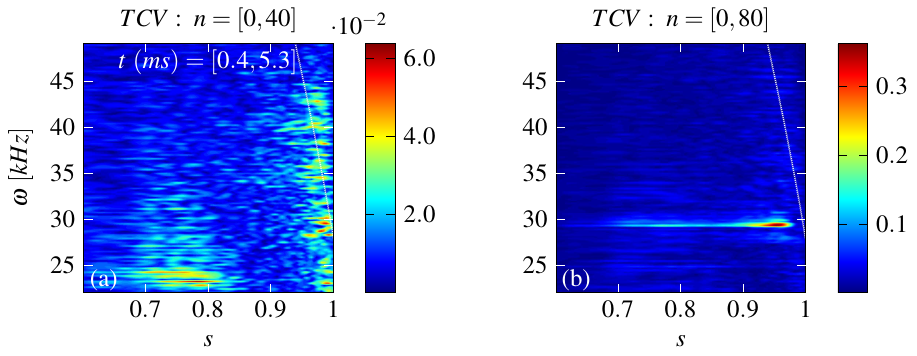}}\\
    \subfloat{\includegraphics[width=0.99\textwidth]{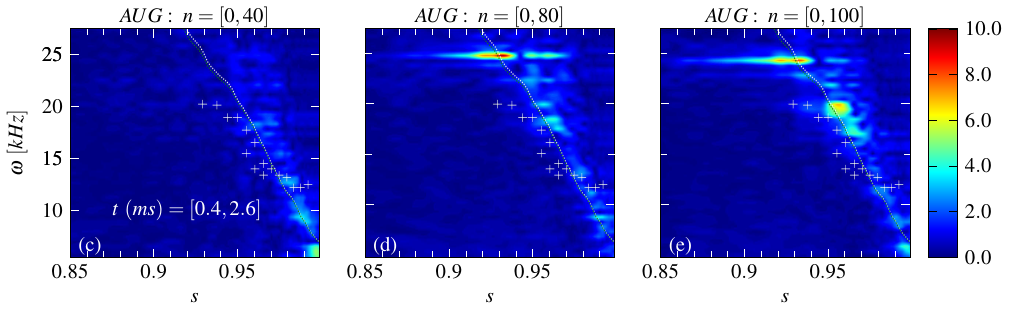}}
    \caption{\label{fig:orig-erbar-fft}
        Frequency spectra of zonal electric field in the TCV (upper row) and the AUG (bottom row) configurations in simulations with different ITG spectra. 
        The global ZS spectra in the TCV are calculated in the time interval $t (ms) = [0.4, 5.3]$, while it is $t (ms) = [0.4, 2.6]$ for the AUG.
        The white dashed lines indicate the linear analytical GAM frequency~\cite{Gao10}. 
        The dashed green lines in the AUG plots depict GAM frequency estimation from linear GK simulations~\cite{Novikau17}.
        The white crosses show experimental frequencies in the GAM range, in the AUG case~\cite{Conway08}.
    }
\end{figure}

The $ExB$ velocity drift, which is proportional to $\ybs{\nabla}\overline\Phi$, is responsible for the global ZS oscillations due to the magnetic field variation on the flux surfaces. 
That is why, it is reasonable to investigate time dynamics of the zonal electric field (instead of the zonal potential $\overline\Phi(t)$) to calculate the GAM frequency spectrum.
There is also a numerical reason why it is necessary to use the zonal electric field for the GAM analysis instead of the potential.
Since the GAMs oscillate close to the edge, they affect the imposed Dirichlet boundary conditions.
To restore them, at every time step the code shifts the zonal potential in the whole radial domain by the non-zero value that the potential has at the edge.
This numerical procedure creates an artificial global oscillation of the electric potential.
However, since the shift is constant in the radial domain, one can eliminate it by analyzing the radial electric field.

Time evolution of zonal electric field generated by different ITG spectra (with $n_{max} = 40$ and $n_{max} = 80$) is shown in Fig.~\ref{fig:orig-erbar-st}.
One can clearly that radially propagated stripes, which oscillate with the same frequency in a wide radial domain, are formed only if one takes into account ITG modes with high toroidal numbers (Figs.~\ref{fig:orig-erbar-st}b and~\ref{fig:orig-erbar-st}d).
The Fourier analysis of these signals (Fig.~\ref{fig:orig-erbar-fft}) confirms the tendency that is observed in both TCV and AUG plasmas.

%% file: antenna.tex
\section{TCV and AUG simulations with antenna}
\label{sec:tcv-aug-antenna}

\begin{figure}[!t]  
    \includegraphics[width=0.99\textwidth]{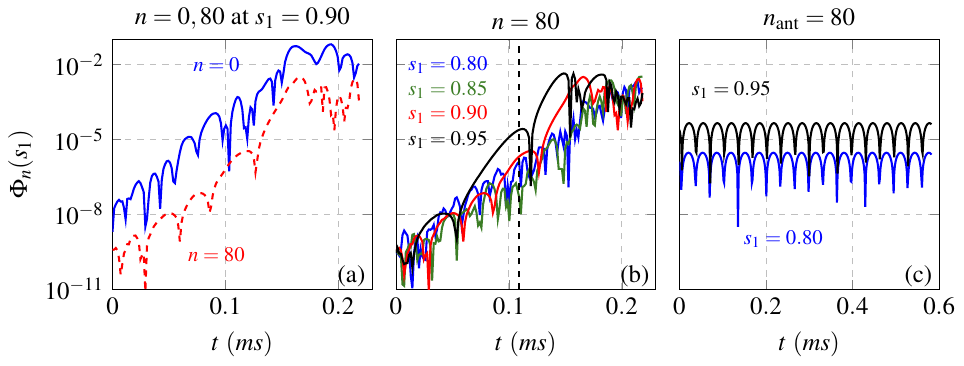}
    \caption{\label{fig:tcv-n-t}
        (a): Temporal evolution of the zonal potential and high-$n$ ITG mode in the TCV simulation with the broad range of toroidal modes $n = [0,80]$.
        (b): The $n = 80$ potential at different radial points in the same TCV simulation. 
        The vertical black dashed line indicates the time instant $t = 0.109$ at which the antenna spatial profile is taken for later simulations.
        (c): Temporal evolution of the $n = 80$ antenna with frequency $\omega(kHz) = 15.31$, whose spatial profile is taken from the TCV simulation with $n = [0,80]$.
    }
\end{figure}
\begin{figure}[!t]  
    \includegraphics[width=0.99\textwidth]{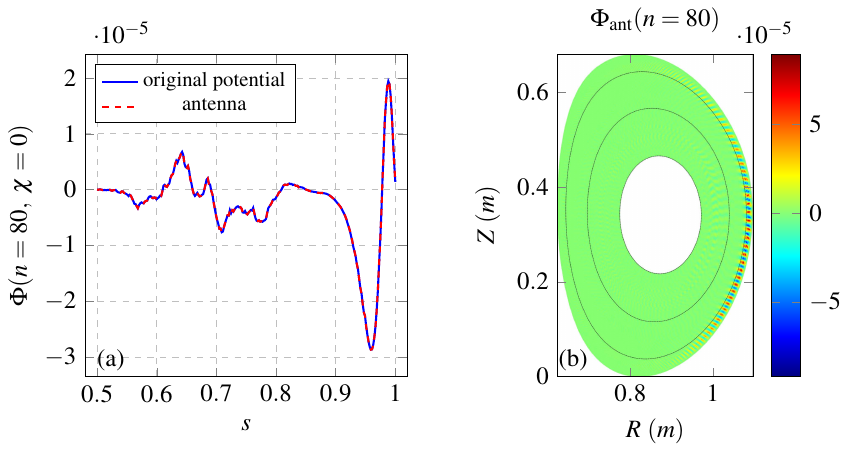}
    \caption{\label{fig:tcv-ant-n80-rad-str}
        (a): Radial structure of the $n = 80$ electric potential at the low field size at $t = 0.109$ in the TCV simulation with $n = [0, 80]$ and the antenna recreated from the this potential.
        (b): Poloidal structure of the antenna.
        The black lines indicate the flux surfaces at $s = 0.50, 0.80, 0.95$, respectively.
    }
\end{figure}

\begin{figure}[!t]  
    \includegraphics[width=0.99\textwidth]{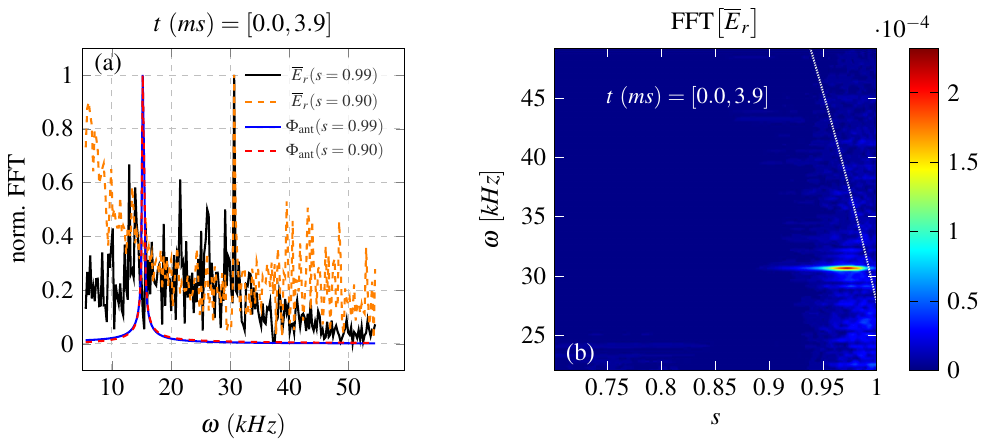}
    \caption{\label{fig:tcv-ant80-res}
        (a): Frequency spectra of the $n = 80$ antenna and the zonal mode generated by the antenna at different radial points in the TCV configuration. 
        (b): Radial distribution of the frequency spectrum of the zonal electric field generated by the $n = 80$ antenna. 
        The white line indicates the linear analytical estimation~\cite{Gao10} of the GAM frequency.
        In both plots, the frequency spectra are calculated in $t (ms) = [0.0, 3.9]$ time interval.
    }
\end{figure}

In the performed computations, the ZSs have been excited by the non-zonal (turbulent) field that naturally arises due to the presence of the plasma temperature gradient.
Moreover, the zonal modes generated by drift instabilities change the amplitude and space structure of initially excited ITG modes.
To simplify the system and concentrate only on the ZS excitation, an ITG antenna can be applied to the same plasma configuration instead of modeling the full ITG dynamics.
In such a way, one can keep the ITG space structure constant during the whole nonlinear simulation by eliminating the back reaction of the zonal modes on the turbulence.
Apart from that, by using the antenna, one has more freedom to control the ZS formation by varying the antenna parameters such as its frequency and toroidal structure.
As it has been shown in Sec.~\ref{sec:tcv-aug-original}, the global ZSs are developed due to the high $n$-modes.
Because of this, the electric potential with $n = 80$ is used here as the ITG antenna.
Its space structure at $t(ms) = 0.109$, Fig.~\ref{fig:tcv-n-t}, is taken from the TCV simulation with the broad toroidal spectrum $n = [0, 80]$, and can be found in Fig.~\ref{fig:tcv-ant-n80-rad-str}.
One can see that the $n = 80$ potential is localized just at the edge, where the zonal has the highest amplitude according to Fig.~\ref{fig:orig-erbar-fft}b.
The mode frequency is estimated to be $\omega(kHz) = 15.31$ within the pre-saturation domain ($t[ms]\sim [0.04, 0.17]$) at $s = 0.90$. 
In general, frequency of any toroidal mode may vary in radial direction.
However, the ITG antenna has constant-in-space frequency (Fig.~\ref{fig:tcv-n-t}c and~\ref{fig:tcv-ant80-res}a).

The created $n = 80$ antenna is applied to the same TCV configuration. 
However, in the simulation with antenna, all modes with non-zero toroidal numbers $n$ are filtered out.
As a result, we have only the $n = 0$ self-consistent field and the $n = 80$ ITG antenna that has influence only on the particle characteristics.
The results of this simulation are shown in Fig.~\ref{fig:tcv-ant80-res}.
The $n = 80$ antenna generates a zonal mode with the frequency $\omega(kHz) \sim 30$ 
(the black and orange lines in Fig.~\ref{fig:tcv-ant80-res}a), which is double of the imposed antenna frequency (red and blue lines). The frequency of the generate zonal mode is close to that with the broad ITG spectrum (compare with Fig.~\ref{fig:orig-erbar-fft}b).
As seen from Fig.~\ref{fig:tcv-ant80-res}b, the excited zonal mode is radially elongated and lies within the radial interval $s\sim [0.90, 1.0]$, while in the TCV simulation with the broad spectrum of toroidal modes the zonal mode spreads up to $s \sim 0.70$.

\begin{figure}[!t]  
    \subfloat{\includegraphics[width=0.99\textwidth]{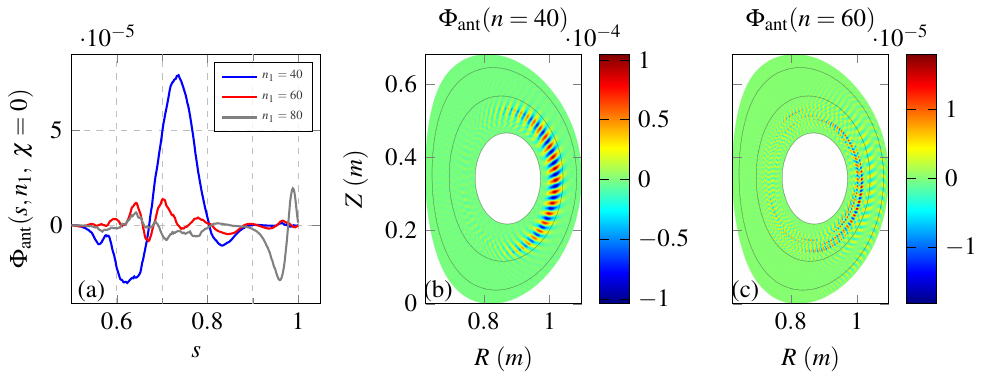}}\\
    \subfloat{\includegraphics[width=0.99\textwidth]{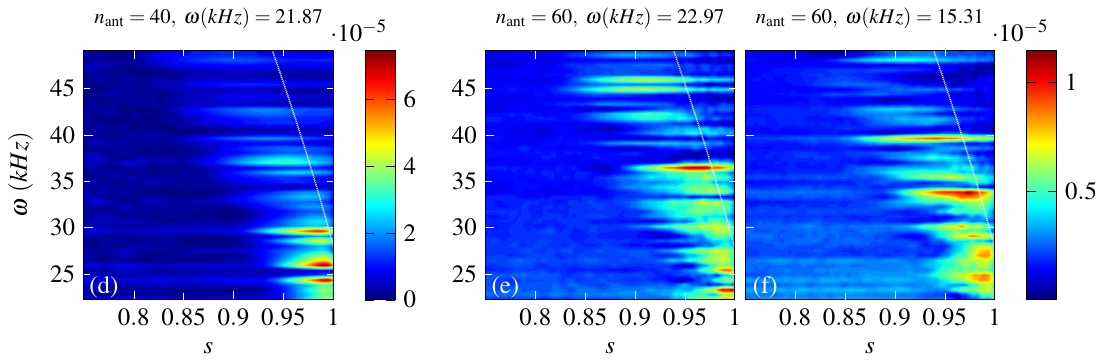}}
    \caption{\label{fig:tcv-a406080}
        (a): Radial structures of the $n = 40$, $n = 60$ and $n = 80$ antennae taken from the TCV simulation with the toroidal spectrum $n = [0,80]$ at $t = 0.109$.
        (b): Poloidal structure of the $n = 40$ antenna.
        (c): Poloidal structure of the $n = 60$ antenna.
        The black lines in plots (b) and (c) indicate the flux surfaces at $s = 0.50, 0.80, 0.95$, respectively.
        (d): Frequency spectrum of the zonal electric field excited by the $n = 40$ antenna with the frequency $\omega(kHz) = 21.87$.
        (e): Frequency spectrum of the zonal mode excited by the $n = 60$ antenna with the frequency $\omega(kHz) = 22.97$.
        (f): Frequency spectrum of the zonal mode excited by the $n = 60$ antenna with the frequency $\omega(kHz) = 15.31$, which is the frequency of the $n = 80$ antenna.
        The Fourier spectra in plots (d), (e), and (f) are calculated in the time interval $t(ms) = [0.0, 1.9]$.
    }
\end{figure}

\begin{figure}[!t]  
    \includegraphics[width=0.99\textwidth]{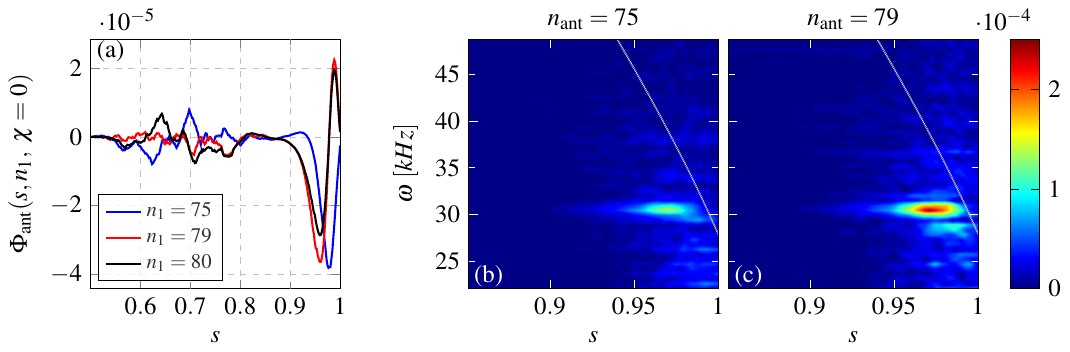}
    \caption{\label{fig:tcv-a7579}
        (a): Radial structure of the antennae with different toroidal mode numbers taken from the TCV simulation with the toroidal spectrum $n = [0,80]$ at $t = 0.109$. 
        Frequency spectra of the zonal mode generated by the $n = 75$ and $n = 79$ antennae are shown in plots (b) and (c), respectively.
        The antennae have the same frequency, $\omega_{ant}(kHz) = 15.31$.
        The Fourier spectra are calculated in the time interval $t = [0.0, 1.9]$.
    }
\end{figure}

The amplitude of the excited global structure is significantly higher than that of the continuum GAM branch, which is similar to what is observed in the simulations with broad ITG spectra, Fig.~\ref{fig:orig-erbar-fft}b.
By applying antennae with smaller toroidal numbers, one can generate the continuum branch, Fig.~\ref{fig:tcv-a406080}.
Although the $n = 40$ and $n=60$ antennae are localized significantly closer to the plasma core, they still give rise to the ZS at the edge, but with much smaller amplitudes.
These antennae have higher frequencies than that of the $n = 80$ mode.
However, even if we use the same frequency for the $n = 60$ and $n = 80$ antennae, the $n = 60$ still does not produce the global zonal structure at $\omega[kHz] = 30$ (Fig.~\ref{fig:tcv-a406080}f).
On the other hand, by keeping the same antenna's frequency and just by slightly varying the toroidal mode number near $n = 80$, Fig.~\ref{fig:tcv-a7579}, one can still generate a global ZS.
The position of the global ZS does not change, but the zonal signal degrades when one shifts towards lower toroidal numbers.

\begin{figure}[!t]  
    \subfloat{\includegraphics[width=0.70\textwidth]{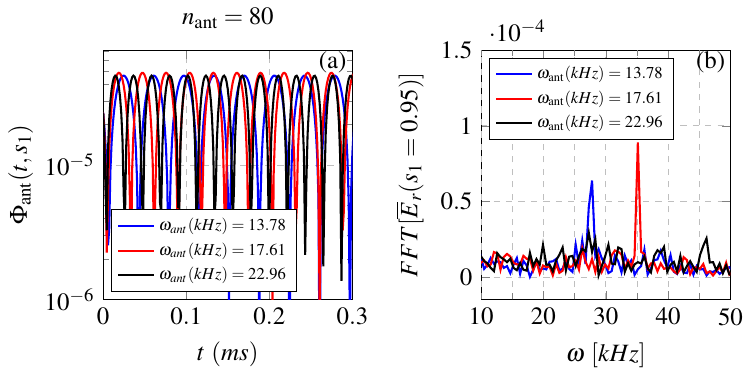}}\\
    \subfloat{\includegraphics[width=0.99\textwidth]{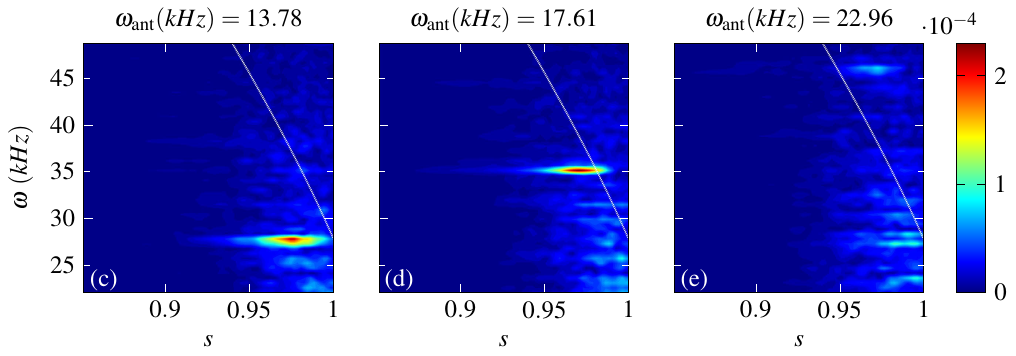}}
    \caption{\label{fig:tcv-a80-diffw}
        (a): Temporal evolution of the $n = 80$ antenna with different frequencies.
        (b): The frequency spectra of the zonal electric field at $s_1 = 0.95$ generated by the $n = 80$ antennae.
        (c), (d), (f): Spatial distribution of the frequency spectra of the zonal electric field generated the $n = 80$ antenna with various driving frequencies. 
        The Fourier spectra are calculated in the time interval $t(ms) = [0.0, 1.9]$.
    }
\end{figure}

Finally, keeping the same toroidal mode $n = 80$ and antenna's amplitude, Fig.~\ref{fig:tcv-a80-diffw}a, one can vary the antenna frequency.
As seen from Fig.~\ref{fig:tcv-a80-diffw}b, the antenna's frequency is one of the key parameters that determines the resulting frequency of the generated zonal signal.
The common tendency is that the resulting zonal frequency is double of that of the imposed antenna:
\begin{equation}
\omega_{\rm zonal} = 2 \omega_{\rm ant}.
\end{equation}
On the other hand, if one strongly shifts the antenna's frequency from its initial value, the amplitude of the resulting zonal mode significantly decreases.
Thus, both spatial and temporal parameters of toroidal are important for successful generation of global ZS.

%% file: Conclusions.tex
\section{Conclusions}
\label{sec:conclusions}

Tokamak plasmas show the presence of turbulence, due to spatial gradients of temperature and density. Turbulence generates zonal, i.e. axisymmetric, structures (ZSs) of the radial electric field. 
ZSs play an important role in the turbulence saturation, with a predator-prey mechanism. 
As turbulence is responsible for the anomalous transport of heat and particles from the tokamak core to the edge, we need to understand the turbulence formation and saturation. 
Therefore, a theoretical model of ZSs is also necessary to achieve.

Zonal structures, and their associated poloidal flows are present with a very low frequency (zero-frequency-zonal-flow) and with finite frequency, the geodesic acoustic modes, GAMs. 
The GAM frequency is of the order of the sound speed divided by the tokamak major radius. 
As the sound speed is a function of the plasma temperature, and as the temperature typically decreases towards the edge, where the GAMs are mostly observed, we can understand that most of the times, the GAM frequency changes in a "continuous" way moving in radius towards the edge. 
This behaviour is called the GAM continuum~\cite{Zonca08}.

Interestingly, ZSs with a radially extended structure and a coherent (i.e. constant in space) frequency have also been observed experimentally~\cite{Huang18}. With the goal of understanding their origin, we have started a linear analysis with the gyrokinetic particle-in-cell code ORB5~\cite{Lanti19}. 
The first analysis has been carried out using linear simulations, approaching realistic experimental conditions as closely as possible~\cite{Novikau17}. 
Although several regimes have been investigated, only GAMs with a continuum radial structure have been observed, leading to the conjecture that the origin of global ZS must be nonlinear, i.e. due to their generation by turbulence. 
A similar conjecture has also been proposed in Ref.~\onlinecite{Merlo18}.

In this paper, we identify the generation mechanism with a two-step process. 
First, we investigate the generation of global ZSs by means of self-consistent numerical simulations, where the global ZSs are observed in the presence of ITG turbulence. 
Second, we reproduce the generation of global ZSs with artificially generated fields, implemented in ORB5 by means of an antenna module. 
The first step is important as it helps identifying the ITGs which are responsible for the global ZS generation. 
They are found to be the ITGs at the high-$n$ part of the spectrum, with $n$ being the toroidal mode number. 
Their spatial and temporal signatures are measured and employed for the antenna in the subsequent step. 
With this process, we are able to reproduce global ZSs in TCV and AUG.

It is interesting to see that one can excite a global ZS with an antenna with a single toroidal mode number, $n=80$ in our selected TCV case. 
This means that this is not a statistical phenomenon due to the presence of many ITG modes forming the turbulence. 
We have also varied the antenna frequency: with this test, we observe that the frequency of the global ZS is twice the antenna frequency. 
This phenomenon can be interpreted as a three-wave coupling between the antenna and a sideband with $n=-80$, rotating at a negative frequency. 
Since the ITG frequency decreases with decreasing $n$, approaching zero as $n \rightarrow 0$, it follows that ITGs with $n$ below a certain threshold cannot drive global ZS, as twice the ITG frequency would lie far outside the typical GAM frequency range.
As next steps, we would like to investigate the propagation of these global ZSs. 
In fact, in Ref.~\onlinecite{Villard19}, simulations on TCV with ORB5 have shown that these are propagating structures. 
Recent studies also demonstrate the importance of trapped electron modes (TEM)~\cite{Feng25, Albert25} in the generation of GAMs.
Modeling this generation mechanism requires the inclusion of kinetic electron effects.
We leave this study to a dedicated publication.